# EMO100DB: AN OPEN DATASET OF IMPROVISED SONGS WITH EMOTION DATA

**Daeun Hwang (a), Saebyul Park (b)**

(a) Yonsei University, South Korea, hdaeun9.8@yonsei.ac.kr,
(b) Yonsei University, South Korea, saebyul_park@yonsei.ac.kr

In this study, we introduce Emo100DB: a dataset consisting of improvised songs that were recorded and transcribed with emotion data based on Russell's circumplex model of emotion. The dataset was developed by collecting improvised songs that consist of melody, lyrics, and an instrumental accompaniment played, sung, and recorded by 20 young adults. Before recording each song, the participants were asked to report their emotional state, with the axes representing arousal and valence based on Russell's circumplex model of emotions. The dataset is organized into four emotion quadrants, and it includes the lyrics text and MIDI file of the melody extracted from the participant recordings, along with the original audio in WAV format. By providing an integrated composition of data and analysis, this study aims to offer a comprehensive dataset that allows for a diverse exploration of the relationship between music and emotion.

## 1. Introduction

Music is oftentimes referred to as a "language of emotion" (Pratt, 1950). Although there has been a significant amount of research on music emotion recognition (MER) in recent years, the majority of the existing studies focus on audio and lyric data. Existing datasets have used various approaches to analyzing emotions, with the most common form being audio files of music annotated with valence and arousal, derived from Russell's circumplex model of emotion (Malheiro et al., 2018; Ricardo, 2017; Panda et al., 2018a; Panda et al., 2018b; Malheiro et al., 2016; Chen et al., 2015). These datasets have been widely used in research on music emotion recognition, and have contributed to the development of various algorithms and systems for detecting and analyzing emotions in music. However, there are also several challenges and limitations associated with the use of these datasets, such as the variability of annotated emotion labels, the limited size and diversity of the datasets, and the complexity of integrating multiple modalities of data. Consequently, ongoing research is focused on the development of new datasets and techniques to enhance the accuracy and reliability of these systems, with multimodal approaches demonstrating higher accuracy (Panda et al., 2018; Panda et al., 2020). A notable observation here is that most research on the MER dataset primarily involved annotators who listened to the music and labeled emotions for each song, resulting in limited datasets that offer emotions annotated from the perspective of songwriters rather than listeners. Additionally, there are few datasets available for analyzing multiple aspects due to the prevalence of audio files.

Therefore, we present Emo100DB, a dataset comprising 100 improvised songs that have been recorded, transcribed, and annotated for music emotion recognition and analysis. This dataset contains lyrics, melody data (in MIDI format), audio data with accompaniment (in WAV format), and emotional annotations corresponding to each data. By integrating a variety of data types and conducting thorough analysis, we aim to provide a comprehensive dataset in order to examine music and emotion from various angles.

## 2. Methods

### 2.1. Participants

To gather the data, we recruited participants who are young adults (age 18-25) and are intermediate or higher level of playing at least one instrument. A total of 29 people participated, out of which 20 successfully completed the assigned tasks, resulting in a final participant sample size of 20 (M_age: 21.7, SD_age: 2.04, 13 females and 7 males). Among the participants, 19 individuals from diverse cultural backgrounds were included. As for their experience of playing the instrument that they used to participate in this experiment, the participants had an average of 10.76 years (SD = 6.73), ranging from 3 months to 20 years of experience.

### 2.2. Materials

We used Russell's circumplex model as a framework for participants to annotate their emotions on arousal and valence scales. Russell's circumplex model illustrated by Du et al., and space for comments on their emotions were provided to the participants (Figure 1). For the post-test questionnaire, we asked questions regarding the following aspects: participants' daily recording of emotions, their understanding of their own emotions, their expression of emotions through lyrics, instruments, and melodies, the potential impact of participating in this study and improvising song on their understanding of emotions, their previous experience with composing songs and playing instruments, and an open section for additional comments or thoughts.

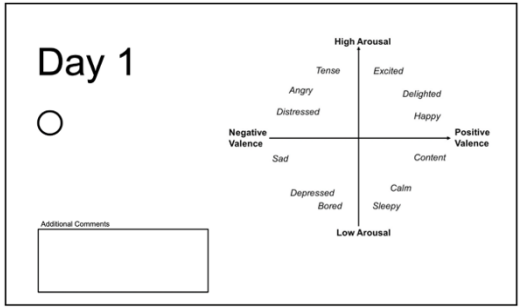

**Figure 1**. Model used to record emotions

### 2.3. Procedure

Participants were asked to improvise a song and record it for five days. The songs had to include a melody, lyrics, and an instrumental accompaniment. Before recording a song, they had to mark their feelings on Russell's circumplex model of emotion. They were requested to record their emotions and songs corresponding to those emotions for five chosen days within a period of 3 weeks. The recording days were not required to be consecutive to enhance participant motivation and avoid any sense of obligation. This approach allowed us to observe what types of emotions the participants recorded the most often. Most participants recorded themselves singing and playing the instrument at the same time, while some also recorded the instrument first and then sang on top of it. They could use any instruments they wanted. Participants submitted their files directly, and a unique participant number was assigned to each file in order to ensure anonymity. Participants were also asked to avoid marking neutral emotions in order to clarify which axis each emotion would fall onto. After recording for 5 days, participants were also asked to fill out the final questionnaire, which consisted of questions asking for basic demographic information and about the overall experience.

### 2.4. Data Analysis

After the dataset was built, we examined the songs with their melodic data for empirical research. Word2vec embedding on the pitch interval was used to analyze data corresponding to each of the four quadrants of emotions. We used the Gensim library[1] to train a Word2Vec model. Pitch intervals were extracted from the MIDI files and utilized to train a Word2Vec model, generating vector embeddings for each distinct pitch interval by capturing their occurrence patterns within the dataset. To observe the linguistic characteristics of each emotion, we used WordCloud and extracted sentiment scores through the transformer library[2] (Wolf et al., 2020). The t-distributed stochastic neighbor embedding (t-SNE) visualization was also used to check if any notable characteristics were observed among different emotions. We used the scikit-learn library to apply t-SNE to the vector representations in two dimensions obtained from the Word2Vec model, with the perplexity parameter of 40.

## 3. Results

### 3.1. Emo100DB

This research has resulted in the creation of a publicly available dataset comprising 100 songs. The dataset includes recorded audio in audio format, transcribed symbolic format for the melody data, lyrics, and corresponding emotions, with a distribution of 29 songs in Quadrant 1 (Q1), 15 songs in Quadrant 2 (Q2), 20 songs in Quadrant 3 (Q3), and 36 songs in Quadrant 4 (Q4). This dataset is publicly open.[3] Melody data is provided in MIDI format. Songs recorded by the participants are provided in WAV format as well as in the normalized format that was trimmed into 15 seconds each. The lyrics of each song are organized in an Excel file (xls format).

### 3.2. Data Analysis Results

The analysis of melodic data using t-SNE visualization (Figure 2) revealed distinct clusters representing different emotions. Notably, we observed that quadrant 1 data was scattered the most widely while quadrant 3 data was condensed closely.

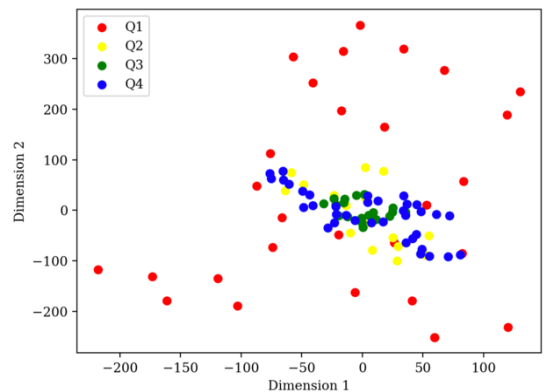

**Figure 2**. t-SNE on melodic data

When applying t-SNE to lyric data (Figure 3), with quadrant 1 exhibiting wide scattering and quadrant 2 showing relatively more condensed clusters.

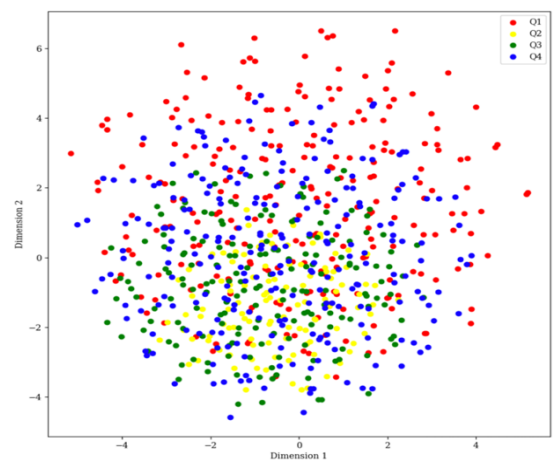

**Figure 3**. t-SNE on lyric data

---

[1] https://pypi.org/project/gensim/
[2] https://pypi.org/project/transformers/
[3] https://github.com/hdaeun98/emo100db

When observed with sentiment analysis on lyrics using the transformer library, the average scores of each emotion showed a quite distinguishable difference, especially for emotions with different valence values. When observed through a density plot, as shown in Figure 4, quadrant 2 showed the most data on negative sentiment scores and the least positive scores. In contrast, quadrant 1 showed the most data with positive sentiment scores and the least negative scores.

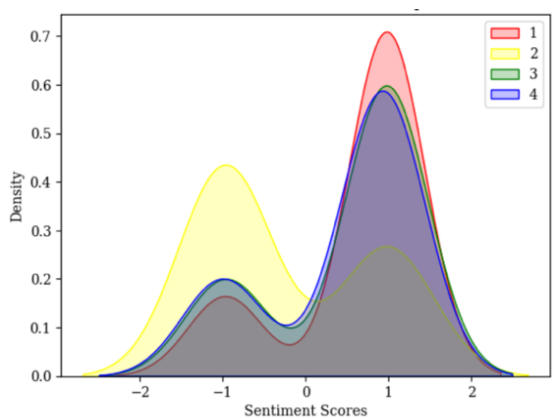

**Figure 4**. Sentiment scores of four emotion quadrants

When the average values of sentiment scores on lyrics for each of the four quadrants were extracted, some quadrants showed distinct differences (Figure 5). Positive sentiment scores, labeled with the color blue in the graph, indicate that the lyrics convey predominantly positive or favorable emotions. Negative sentiment scores, conversely, indicate that the lyrics predominantly convey negative or unfavorable emotions. For both emotions, higher scores would indicate a stronger presence of corresponding emotion within the lyrics. Quadrant 1 and quadrant 2 showed the most remarkable differences, since quadrant 1 showed the highest average value around 0.72, while quadrant 2 was negative around -0.23. Quadrant 3 and quadrant 4 showed a subtle difference, with values of 0.49 and 0.32, respectively.

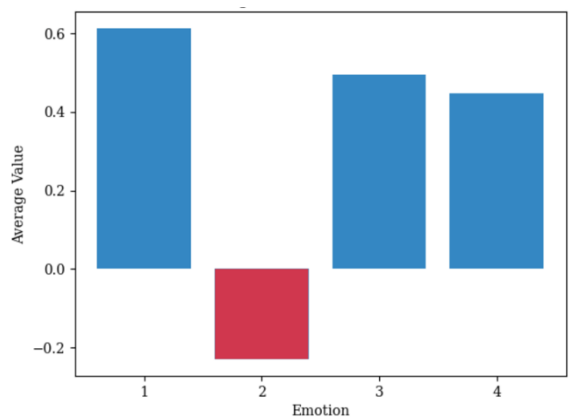

**Figure 5**. Average of sentiment scores for each four quadrants

With WordCloud, we analyzed what kind of words in lyrics each quadrant could be represented with (Figure 6). The largest words shown for each emotion demonstrated a distinctive representation.

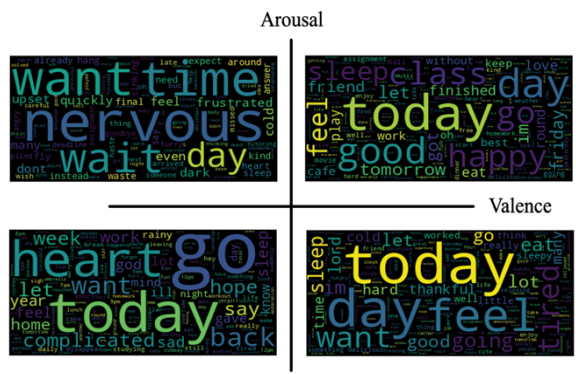

**Figure 6**. WordCloud for lyrics of emotion for each quadrant

WordCloud for quadrants 2 and 3 which both had negative valence showed clearer words of negative emotions, such as 'upset', 'nervous', 'complicated', and 'sad.' Quadrant 1 with positive valence and positive arousal showed evidently positive words such as 'happy' and 'good.' Meanwhile, quadrant 4 showed a mixture of positive and negative words, such as 'good' and 'thankful', but also 'tired' and 'hard'.

### 3.3. Post-Test Questionnaire

On the post-test questionnaire questions asking about the experience of improvising songs and recording emotions, participants expressed that this activity helped them understand their emotions better (M = 4, SD = .83). Participants' scores implied that they expressed their feelings the best in order of lyrics, music, and melody. No participant recorded the same emotion more than 3 times during 5 days of recording. It was also observed that participants recorded the most in the order of quadrants 4, 1, 3, and 2, which may signify that young adults tend to record their emotions in songs when they are feeling emotions that correspond to low arousal and low valence.

### 4. Discussion

In this study, we have constructed Emo100DB, an integrated dataset for research on music emotion recognition. Emo100DB consists of improvised songs recorded and transcribed with emotion data based on Russell's circumplex model. The dataset is categorized into four emotion quadrants and includes lyrics text, MIDI files of the melodies from participant recordings, and the original audio in WAV format. This dataset provides a valuable resource for exploring the relationship between improvised songs and emotional expression. Its inclusion of lyrics text, MIDI files, and original audio would expand possibilities for various analyses and applications as well.

Through our analysis of the collected data using various models and visualizations, we have observed distinct differences in the lyrics and melodies of songs with different emotions. Additionally, we have utilized Word2Vec embeddings and sentiment analysis to analyze the linguistic characteristics of melodies corresponding to each emotion, expanding the possibilities of using natural language processing (NLP) to analyze music as an emotional language.

We recognize the limitations of our study, primarily the small number of participants and the limited amount of data collected. Future studies could include a wider age range and a larger number of participants to verify more detailed characteristics and analyze improvised songs with emotion data. Additionally, incorporating emotional annotations from listeners would validate potential differences between listener and songwriter perceptions of music, contributing to significant discoveries in the interplay between music and emotion.